\documentclass[12pt]{article}

\usepackage{epsfig}
\usepackage{cite}
\usepackage{amssymb}
\begin{document}
\title{Cooperative molecular motors moving back and forth}
\author{David Gillo$^{1\dagger}$, Barak Gur$^{2\dagger}$,\\
Anne Bernheim-Groswasser$^1$ and  Oded Farago$^3$ \\
$^1$Department of Chemical Engineering, $^2$Department of 
Physics, \\
$^3$Department of Biomedical Engineering \\ 
Ben Gurion University, Be'er Sheva 84105, Israel}
\maketitle
\begin{abstract}
We use a two-state ratchet model to study the cooperative
bidirectional motion of molecular motors on cytoskeletal tracks with
randomly alternating polarities. Our model is based on a previously
proposed model [Badoual et al., {\em Proc. Natl. Acad. Sci. USA} {\bf
99}, 6696 (2002)] for collective motor dynamics and, in addition,
takes into account the cooperativity effect arising from the elastic
tension that develops in the cytoskeletal track due to the joint
action of the walking motors. We show, both computationally and
analytically, that this additional cooperativity effect leads to a
dramatic reduction in the characteristic reversal time of the
bidirectional motion, especially in systems with a large number of
motors.  We also find that bidirectional motion takes place only on
(almost) a-polar tracks, while on even slightly polar tracks the
motion is unidirectional. We argue that the origin of these
observations is the sensitive dependence of the cooperative dynamics
on the difference between the number of motors typically working in
and against the instantaneous direction of motion.
\end{abstract}
\vspace{4cm}
$^\dagger$ Authors with equal contributions

\newpage

\section{Introduction}
\label{sec:introduction}

Many cellular processes such as cell motility and mitosis require the
cooperative work of many motors in order to preserve continuous motion
and force generation \cite{vale}. Muscle contraction, for example,
involves the simultaneous action of hundreds of myosin II motors
pulling on attached actin filaments and causing them to slide against
each other \cite{geeves}.  Groups of myosin II motors are also
responsible for the contraction of the contractile ring during
cytokinesis \cite{feierbach}. In certain biological systems,
cooperative behavior of molecular motors produces oscillatory
motion. In some insects, for instance, cooperative behavior of
molecular motors leads to oscillations of the flight muscles
\cite{machine}. Another example is the oscillatory motion of axonemal
cilia and flagella, which is believed to be generated by a large
number of interacting dynein motors \cite{camalet,brokaw}. Finally,
cooperative action of motors is required for the extraction of
membrane tubes from vesicles \cite{koster,leduc}

One of the more interesting outcomes of cooperative action of
molecular motors is their ability to induce bidirectional
motion. ``Back and forth'' dynamic has been observed in various
motility assays including: (i) myosin II motors walking on actin
tracks with randomly alternating polarities \cite{gilboa}, (ii) NK11
(kinesin related Ncd mutants which individually exhibit random motion
with no preferred directionality) moving on microtubules (MTs)
\cite{endow}, (iii) mixed population of plus-end (kinesin-5 KLP61F)
and minus-end (Ncd) driven motors acting on MTs \cite{tao}, and (iv)
myosin II motors walking on actin filaments in the presence of
external stalling forces \cite{riveline}. Reversible transport of
organelles through the combined action of kinesin II, dynein, and
myosin V has been also observed in {\em Xenopus}\/ melanophores
\cite{gross}. In the latter example, the kinesin and dynein move the
organelle in opposite directions along MTs, while the myosin motors
(which take the organelle on occasional ``detours'' along the actin
filaments) function as ``molecular ratchets'', controlling the
directionality of the movement along the MT transport system. From a
theoretical point of view, cooperative dynamics of molecular motors
and, in particular, bidirectional movement, have been investigated
using several distinct models. These models include: (i) lattice and
continuum asymmetric exclusion models
\cite{chowdhury,frey,campas,muhuri,chowdhury2,lichtenthaler,santen},
(ii) ratchet models of {\em interacting}\/ particles moving in the
presence of a periodic potential
\cite{campas,frank1,frank2,badoual,dasilva}, and (iii) the tug-of-war
model which has been recently proposed for describing the transport of
cargo by the action of a few motors
\cite{lipowsky1,lipowsky2,zhang,hexner}. The common theme in these
experimental and theoretical studies is the association of
bidirectionality with the competition between two populations of
motors that work against each other to drive the system in opposite
directions. The occasional reversals of the transport direction
reflect the ``victory'' of one group over the other during the
respective time intervals. The balance of power is shifting between
the two motor parties as a result of stochastic events of binding and
unbinding of motors to the cytoskeletal track. Without going into the
details of the various existing models of cooperative bidirectional
motion, we note that most of them assume that the motors interact
mechanically but act independently, i.e., their binding to and
unbinding from the track are uncorrelated. By further assuming that
the attachment and detachment events of individual motors are
Markovian, the distribution of ``reversal times'' (i.e., the durations
of unidirectional intervals of motion) can be shown to take an
exponential form
\begin{equation}
p(\delta t)=\exp(-\delta t/\tau_{\rm rev}), 
\label{eq:exponential}
\end{equation}
where $\tau_{\rm rev}$ is the characteristic reversal time of the
bidirectional motion.

\begin{figure}
  {\centering \hspace{2cm}\epsfig{file=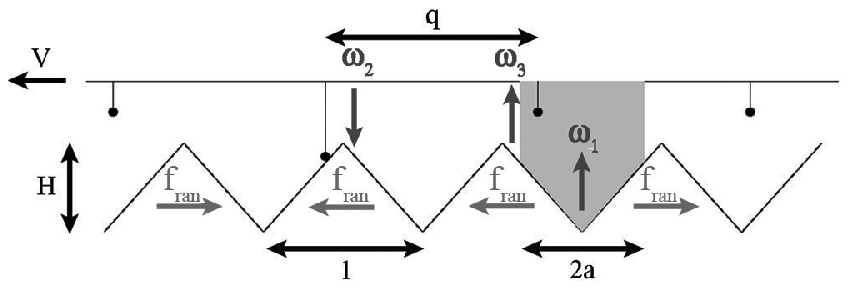,width=14cm}}
\caption{$N$ point particles (representing the motors) are connected
to a rigid rod with equal spacing $q$. The motors interact with the
actin track via a periodic, symmetric, saw-tooth potential with period
$l$ and height $H$. In each periodic unit, there is a random force of
size $f_{\rm ran}$, pointing either to the right or to the left. The
motors are subject to these forces only if connected to the track. The
detachment rate $\omega_1$ is localized in the shaded area of length
$2a<l$, while the attachment rate $\omega_2$ is located outside of
this region. The off rate $\omega_3$ is permitted only outside the
gray shaded area.}
\label{fig:1}
\end{figure}

The magnitude of $\tau_{\rm rev}$ can be taken as a measure for the
degree of cooperativity between the motors. The more cooperative the
motors are, the more persistent is the movement and the longer are the
periods of unidirectional transport. The run lengths (in each
direction) of highly cooperative motors may be of a few microns even
for non-processive motors like myosin II \cite{gilboa}. As noted
above, the mechanical coupling between the motors is sufficient for
the generation of highly cooperative bidirectional motion, even if the
motors attach to/detach from the track in an uncorrelated
fashion. This has been demonstrated theoretically by Badoual {\em et
al.}\/~some years ago \cite{badoual}. A slightly modified version of
this model is illustrated schematically in Fig.~\ref{fig:1}.  The
model considers the one-dimensional motion of a group of $N$ point
particles (representing the motors) connected (mechanically coupled)
to a rigid rod with equal spacing $q$. The cytoskeletal track is
represented by a periodic saw-tooth potential, $U(x)$, with period $l$
and height $H$. The model requires that $q$ is larger than and
incommensurate with $l$. The motors are identical and walk on a track
which is globally a-polar and, thus, does not permit net transport to
the right or left over large time scales. The temporal direction of
motion is determined by the net force generated by all the motors.
The local polarity of the track is represented by an additional force
of size $f_{\rm ran}$ (denoted by a horizontal arrow in each periodic
unit in Fig.~\ref{fig:1}) which, within each unit of the periodic
potential, points to the right or to the left. The globally a-polar
nature of the track is ensured by requiring that the sum of these
random forces vanishes.

The instantaneous force between the track and the motors is given by
the sum of all the forces acting on the individual motors:
\begin{equation}
F_{\rm tot}=\sum_{i=1}^{N}f_i^{\rm motor}=\sum_{i=1}^{N}\left[
-\frac{\partial U\left(x_1+\left(i-1\right)q\right)}{\partial x}+
f_{\rm ran}\left(x_1+\left(i-1\right)q\right)\right]\cdot C_i(t),
\label{eq:ftot}
\end{equation}
where $x_i=x_1+(i-1)q$ is the coordinate of the $i$-th motor. The two
terms in the square brackets represent the forces due to the symmetric
saw-tooth potential and the additional random local forces acting in
each periodic unit. The function $C_i(t)$ takes two possible values, 0
or 1, depending on whether the motor $i$ is detached from or attached
to the track, respectively, at time $t$. The motors change their
binding states (0 - detached; 1 - attached) independently of each
other, according to the following rules: We define an interval of size
$2a<l$ centered around the potential minima (the gray shaded area in
Fig.~\ref{fig:1}). If located in one of these regions, an attached
motor may become detached ($1\rightarrow 0$) with a probability per
unit time $\omega_1$.  Conversely, a detached motor may attach to the
track ($0\rightarrow 1$) with transition rate $\omega_2$ only if
located outside this region of size $2a$.

At each instance, the group velocity of the motors is proportional to the 
total force exerted by the motors (Eq.~\ref{eq:ftot})
\begin{equation}
v(t)=F_{\rm tot}(t)/\lambda,
\label{eq:totalforce}
\end{equation}
where the friction coefficient, $\lambda$, depends mainly on motors
attached to the track and is, therefore, taken proportional to the
number of connected motors, $N_c\leq N$ at time $t$:
$\lambda=\lambda_0 N_c$.  Because the track is globally a-polar, it is
clear that the motors exhibit ``back and forth'' motion with vanishing
mean velocity and displacement. The characteristic time of movement in
each direction, $\tau_{\rm rev}$, may nevertheless be macroscopically
large. The origin of this feature (which reflects the cooperative
character of the motors' action) can be explained as follows: The
stochastic equations of motion of our model system have two identical
(except for sign reversal) steady state solutions corresponding to
right and left motion of the mechanically coupled motors. Each of
these solutions is characterized by $N_c=N\cdot P$ ($P\leq 1$)
connected motors. The connected motors are partitioned to $N_{+}$ and
$N_{-}=N_c-N_{+}\leq N_{+}$ motors that, respectively, support and
object the motion.  Let us define the excess number of motors working
in the direction of the motion as $N\cdot \Delta=N_{+}-N_{-}$, where
$\Delta$ will be termed the ``bias parameter''. Notice that $P$ and
$\Delta$ denote the averages of quantities (which we, respectively,
denote by $P(t)$ and $\Delta(t)$) whose values fluctuate in time due
to the stochastic binding and unbinding of motors. To switch the
direction of motion, $\Delta(t)$ must vanish, and the occurrence
probability of this event, $\Pi(\Delta(t)=0)\sim (\tau_{\rm
rev})^{-1}$.  In section \ref{sec:analytical} we derive an approximate
expression for $\Pi(\Delta(t)=0)$ and show that the mean reversal time
of the bidirectional motion increases exponentially with the size of
the system:
\begin{equation}
\tau_{\rm rev}\sim\left[1-P+\sqrt{P^2-\Delta^2}\right]^{-N}.
\label{eq:taurevexp}
\end{equation}
Thus, for sufficiently large $N$, the characteristic reversal time of
the bidirectional motion becomes macroscopically large. In the
``thermodynamic limit'' ($N\rightarrow \infty$), $\tau_{\rm rev}$
diverges and the motion persists in the direction chosen at random at
the initial time.

The validity of Eq.~\ref{eq:taurevexp} was recently tested using a
motility assay in which myosin II motors drive the motion of globally
a-polar actin bundles \cite{gilboa}. In contrast to the predicted
exponential dependence of $\tau_{\rm rev}$ on the number of working
motors, the experimentally measured reversal times showed no
dependence on $N$. The apparent disagreement between the theoretical
model and the experimental results can be reconciled by noting that
Eq.~\ref{eq:taurevexp} describes exponential growth of $\tau_{\rm
rev}$ with $N$ only when $P$ and $\Delta$ are themselves independent
of $N$. This is indeed the case in the original model presented by
Badoual {\em et al.}\/~\cite{badoual}, where both the on ($\omega_2$)
and off ($\omega_1$) rates do not depend on $N$. In ref.~\cite{gilboa}
we introduced a slightly modified version of Badoual's model, which to
a large extent explains the experimentally observed independence of
$\tau_{\rm rev}$ on $N$. We argued that the origin of this behavior
can be attributed to the tension developed in the actin track due to
the action of the attached myosin II motors. An increase in the number
of attached motors leads to an increase in the mechanical load which,
in turn, leads to an increase in the detachment rate of the motors, as
already suggested in models of muscle contraction
\cite{smith,duke,lan,deville}. But unlike most previous studies where
the myosin conformational energy was calculated, in ref.~\cite{gilboa}
we considered the elastic energy stored in the actin track. Within a
mean field approximation, this energy scales as $E\sim \langle F_{\rm
tot}^2\rangle/k_{\rm sp}$, where $k_{\rm sp}$ is the effective spring
constant of the track and $F_{\rm tot}$ is the total force exerted by
the motors (see Eq.~\ref{eq:totalforce}). The total force is the sum
of $N_c$ random forces working in opposite directions. Therefore, the
mean force $\langle F_{\rm tot}\rangle=0$, while $\langle F_{\rm
tot}^2\rangle$ scales linearly with $N_c$. The spring constant is
inversely proportional to the length of the track, i.e. to the size of
the system and to the total number of motors $N$. We thus conclude
that the mean elastic energy of the actin scales like $E/k_BT\sim
NN_C$, which means that the detachment of a motor ($N_c\rightarrow
N_c-1$) leads, on average, to an energy gain $dE/k_BT=-\alpha N$
($\alpha$ is some dimensionless number). This effect can be
incorporated within the model by introducing an additional off rate,
$\omega_3=\omega_3^0\exp(\alpha N)$, outside the gray shaded area in
Fig.~1 (i.e., around the potential maxima). Thus, in the modified
model, the on and off rates within each unit cell ($-l/2\leq x\leq
+l/2$) of the periodic potential are given by
\begin{equation}
\omega_{\rm\, on} (x)=\left\{\begin{array}{ll}
0 &\ |x|\leq a \\ 
\omega_2 &\ a<|x|\leq l/2 
\end{array}\right.
\label{eq:onrate}
\end{equation}
and 
\begin{equation}
\omega_{\rm\, off} (x)=\left\{\begin{array}{ll}
\omega_1 &\ |x|\leq a \\ 
\omega_3^0\exp(\alpha N) &\ a<|x|\leq l/2. 
\end{array}\right.
\label{eq:offrate}
\end{equation}
The dependence of $\omega_{\rm off}(x)$ on $N$ is another, indirect,
manifestation of cooperativity between the motors which is mediated
through the forces that the motors jointly exert on the actin
track. Notice, however, that the cooperative action of the motors does
not lead (within the modified model) to correlations between
attachment and detachment events of different motors. Moreover, the
attachment/detachment rates do not depend on the number of connected
motors and, thus, are fixed over time. In ref.~\cite{gilboa} we
compared the predictions of the original and modified models for model
parameters corresponding to the myosin II-actin motility assay. In the
former, the reversal times grew exponentially with $N$ from $\tau_{\rm
rev}\sim 1$ sec for $N\sim 1000$, to $\tau_{\rm rev}\sim 10^3$ sec for
$N\sim 3000$. The modified model showed a much better agreement with
the experimental results \cite{gilboa}. Specifically, the reversal
time did not grow exponentially with $N$ but rather showed a weak
maximum around $N\sim 2000$, where $\tau_{\rm rev}\sim 10$ sec.

In this paper we present a more detailed account of the theoretical
model. Several aspects of the model not studied in ref.~\cite{gilboa}
will be discussed in section \ref{sec:computational}, including the
dynamics on non-random and slightly polar tracks. The steady state
solutions of the bidirectional motion are derived analytically in
section \ref{sec:analytical}. We use these solutions to estimate the
reversal times and compare our analytical predictions with the
computational results. We summarize and discuss the results in section
\ref{sec:summary}.

\section{Computational results}
\label{sec:computational}

The model has been presented in details in ref.~\cite{gilboa}, as well
as above in section \ref{sec:introduction}. The model parameters used
in this paper are summarized in Table~\ref{tab:parameters}. The choice
of these values which represent various chemical and physical
parameters of the myosin II-actin system is explained in detail in
ref.~\cite{gilboa}. The model features two new parameters which do not
appear in the original model~\cite{badoual}. The off rate $\omega_3^0$
represents the probability of a motor to detach from the track without
completing a unit step, and its value was estimated in
ref.~\cite{gilboa} by noting that in the absence of an elastic load,
the probability of such an event is 1-2 orders of magnitude smaller
than the complementary probability that the attached motor will
execute the step. The constant $\alpha$ depends on the effective
elastic spring constant of the basic actin unit (monomer) as well as
on the magnitude of the forces that the motors typically apply on the
track (see Eq.~5 in ref.~\cite{gilboa}). The best fit to the
experimental data is achieved with $\alpha=0.0018$ \cite{gilboa},
which gives a weak non-monotonic dependence of $\tau_{\rm rev}$ on
$N$. Here we set $\alpha=0.002$, which gives somewhat poorer agreement
with the experimental results but which highlights the differences
between Badoual's model~\cite{badoual} and the newly proposed model
that takes the elasticity of the actin track into account. As noted
above, the consideration of the elastic properties of the actin and
the cooperative nature of the action of the working motors
considerably improves the results of the original model by replacing
the very strong exponential dependence of $\tau_{\rm rev}$ on $N$ with
a much weaker non-monotonic dependence.

\begin{table}[ht]
\begin{center}
\begin{tabular}{ l  l  }
    parameter & value \\
    \hline
    \hline
    $l$ - period length of the potential & 5 nm \\
    $q$ - spacing between adjacent motors & $(5\pi/12)l\sim 6.54$ nm \\ 
    $2a$ - size of the gray shaded area (see Fig.~7A) & 3.8 nm \\ 
    $2H/l$ - force due to the periodic potential & 5 pN \\
    $f_{\rm ran}$ - random force in each unit cell & 1 pN \\
    $(\omega_1)^{-1}$ {\protect (see Eq.~\ref{eq:offrate})} & 0.5 msec \\
    $(\omega_2)^{-1}$ {\protect (see Eq.~\ref{eq:onrate})} & 33 msec \\
    $(\omega_3^0)^{-1}$ {\protect (see Eq.~\ref{eq:offrate})} & 7500 msec \\
    $\alpha$ {\protect (see Eq.~\ref{eq:offrate})} & 0.002 \\
    $\lambda_0$ - friction coefficient per connected motor & $85\times 10^3$ kg/sec
\end{tabular}
\end{center}
\caption{Values of the model parameters as used in our simulations. }
\label{tab:parameters}
\end{table}

\begin{figure}
  {\centering \hspace{2cm}\epsfig{file=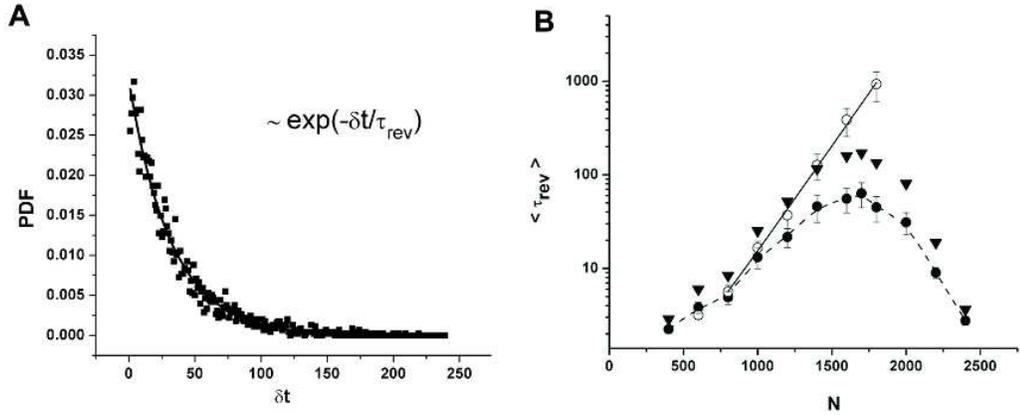,width=14cm}}
\caption{(A) The probability distribution function (PDF) of reversal
times corresponding to one track realization.  The distribution is
fitted by a single exponential decay function - see {\protect
Eq.~\ref{eq:exponential}}. (B) The mean reversal time $\langle
\tau_{\rm rev}\rangle$ as a function of the number of motors $N$.  For
each value of $N$, the calculation of $\langle \tau_{\rm rev}\rangle$
is based on simulations of 40 different track realizations, where the
error bars represent the standard deviation of $\tau_{\rm rev}$
between realizations. The solid and open circles denote the results
corresponding to $\alpha=0.002$ (our model) and $\alpha=0$ (original
model presented in {\protect ref.~\cite{badoual}}), respectively. In
the latter case $\langle \tau_{\rm rev}\rangle$ increases
exponentially with $N$ (as indicated by the solid straight line),
while in the former case $\langle \tau_{\rm rev}\rangle$ exhibits a
non-monotonic behavior (as indicated by the dashed line which serves
as a guide to the eye) and reaches considerably lower values. The
triangles denote the results for periodic tracks whose
reversal times are always larger than those measured on random
tracks.}
\label{fig:2}
\end{figure}

To simulate conditions corresponding to dynamics on a-polar tracks, we
randomly chose the direction of random force (the force representing
the local polarity of the track, see horizontal arrows in
Fig.~\ref{fig:1}) in each unit cell, but discarded the tracks at which
the sum of random forces did not exactly vanish. We computationally
measured characteristic reversal time, $\tau_{\rm rev}$, as a function
of $N$ in the range of $400\leq N\leq 2400$. For each value of $N$, we
generated 40 different realizations of random tracks (each of which
consisting of $M\simeq (q/l)N$ units with periodic boundary
conditions) and simulated the associated dynamics for a total period
of $2\cdot10^5$ seconds. During this period of time we followed the
changes in the direction of motion and calculated the probability
distribution function (PDF) of the reversal times. The characteristic
reversal time corresponding to each random track was extracted by
fitting the PDF to an exponential form (see Eq.~\ref{eq:exponential}),
as demonstrated in Fig.~\ref{fig:2}A.  Fig.~\ref{fig:2}B summarizes
our results, where here for each $N$ the reversal time plotted
(denoted by $\langle \tau_{\rm rev} \rangle$) is the average of
$\tau_{\rm rev}$ calculated for the different track realizations. The
error bars represent the standard deviation of $\tau_{\rm rev}$
between realizations.  The data points depicted in solid circles
correspond to $\alpha= 0.002$, while the open circles correspond to
$\alpha=0$, i.e., to the model originally presented in
ref.~\cite{badoual} where the on and off rates defined in
Eqs.~\ref{eq:onrate} and \ref{eq:offrate} are independent of $N$. As
predicted by Eq.~\ref{eq:taurevexp} and indicated by the straight line
in Fig.~2B, for $\alpha=0$ the mean reversal time $\langle \tau_{\rm
rev} \rangle$ exhibits a very strong exponential dependence on
$N$. Because of this very rapid increase of $\langle \tau_{\rm rev}
\rangle$ with $N$, the reversal times (in the $\alpha=0$ case) could
not be accurately measured for $N>1800$ . Based on the exponential fit
(solid line in Fig.~2B), we estimate that for $N=2400$ the mean
reversal time will be of the order of a few hours.  In contrast, the
calculated $\langle \tau_{\rm rev}\rangle$ corresponding to
$\alpha=0.002$ show a non-monotonic dependence on $N$. The computed
$\langle \tau_{\rm rev}\rangle$ are much smaller in this case, and
fall below 1 minute for all values of $N$. In ref.~\cite{gilboa}, we
used slightly different values of the model parameters than those
given in Table \ref{tab:parameters}. For the model parameters in
ref.~\cite{gilboa}, the variation of $\langle \tau_{\rm rev}\rangle$
with $N$ was even weaker than in Fig.~2B and the largest computed
$\langle \tau_{\rm rev}\rangle\leq 12$ sec. These computational
results were in a very good quantitative agreement with the
experimental results of the in vivo actin-myosin motility assay.

Our simulations reveal surprisingly large variations in the $\tau_{\rm
rev}$ values between random tracks of similar size (see error bars in
Fig.~\ref{fig:2}B). The origin of these variations lies in the fact
that the spacing between motors is larger than the periodicity of the
ratchet potential ($q>l$) and, thus, only $N$ out of $M\simeq (q/l)N$
unit cells are ``occupied'' with motors (which may be either connected
or disconnected) at each instance. Thus, although the track is
perfectly a-polar and contains an equal number of cells with random
forces pointing in both directions, the subset of occupied cells may
have net polarity which constantly changes with time as the motors
move collectively along the track. The direction of the net polarity
of the occupied cells is also the instantaneous preferred direction of
motion. Therefore, the temporal variations in the net polarity must be
correlated with the changes in the directionality of the motion. We
thus expect tracks on which the net polarity changes more frequently
to have smaller $\tau_{\rm rev}$. The effect of net polarity
fluctuations does not occur when the motion takes place on periodic
a-polar tracks, because in this case the equally spaced motors occupy
equal number of cells with left- and right-pointing random
forces. Therefore, the reversal times on periodic a-polar tracks are
expected to be (i) independent of the periodicity of the track, and
(ii) larger than the reversal time on random a-polar tracks. These
predictions are fully corroborated by the results from simulations
with two very distinct periodic tracks - one with period 2 (i.e.,
where the local random force changes its sign every unit cell) and one
with period $M$ (i.e., when the track is divided into two equal
domains of opposite polarities). The results from these two sets of
simulations are denoted by triangles in Fig.~\ref{fig:2}B. The
reversal times of both periodic tracks are nearly indistinguishable
from each other (the differences between them are smaller than the
size of the symbols) and are larger than the reversal times measured
for all the random tracks of similar size.

\begin{figure}
  {\centering \hspace{2cm}\epsfig{file=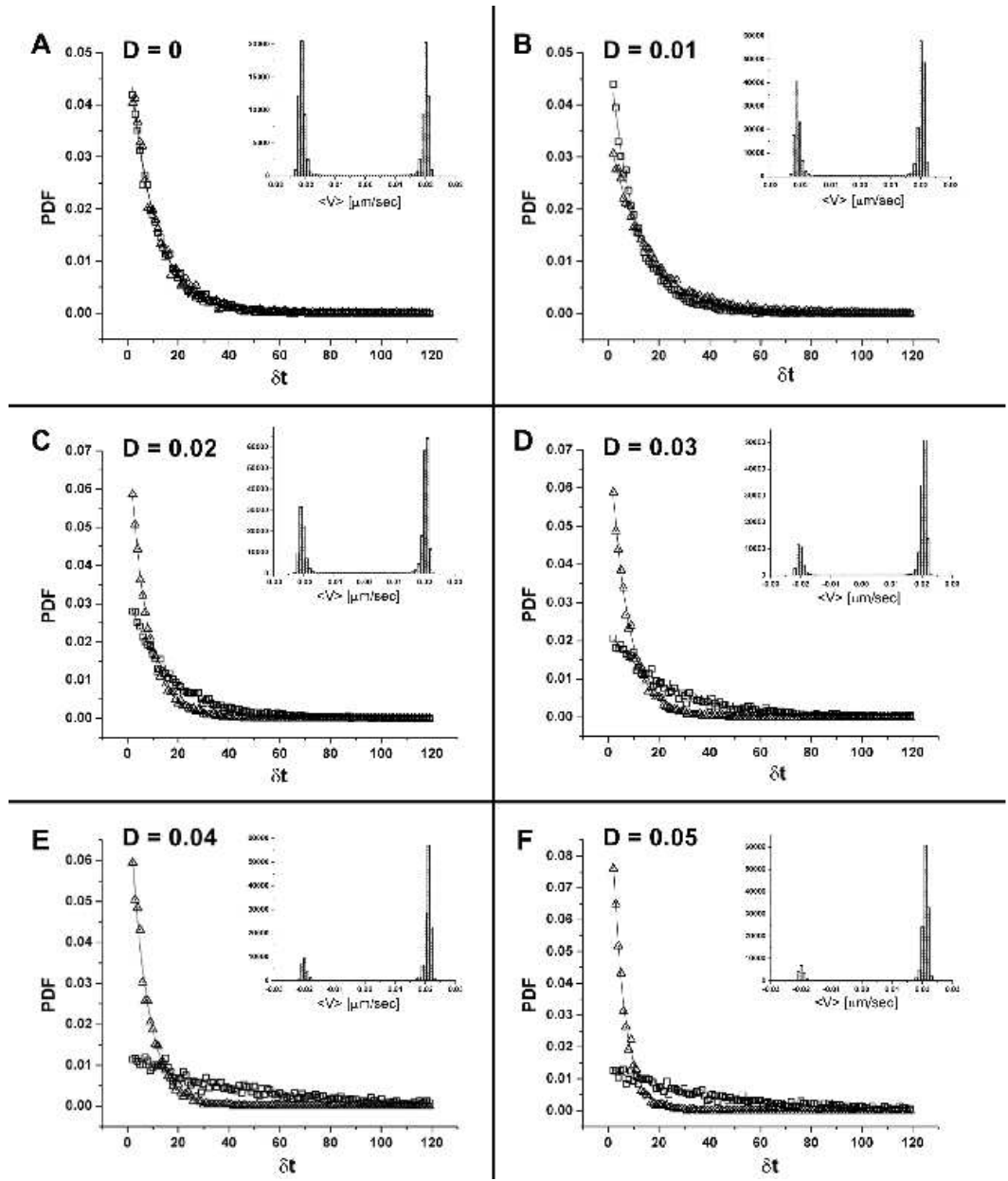,width=14cm}}
\caption{The probability distribution functions (PDF) of reversal
times corresponding to $N=1000$ motors moving on slightly polar
tracks. The variable $D$ denotes the difference between the fraction
of random forces pointing in the favored and disfavored
directions. The motion in the favored and disfavored directions are
analyzed by different PDFs, each of which can be fitted by a single
exponential form but with distinct reversal times (except for $D=0$
where the two PDFs coincide with each other). The insets
show the velocity histogram corresponding to each value of $D$. As $D$
increases, the histograms become less and less symmetric.}
\label{fig:3}
\end{figure}

\begin{figure}
  {\centering \hspace{2cm}\epsfig{file=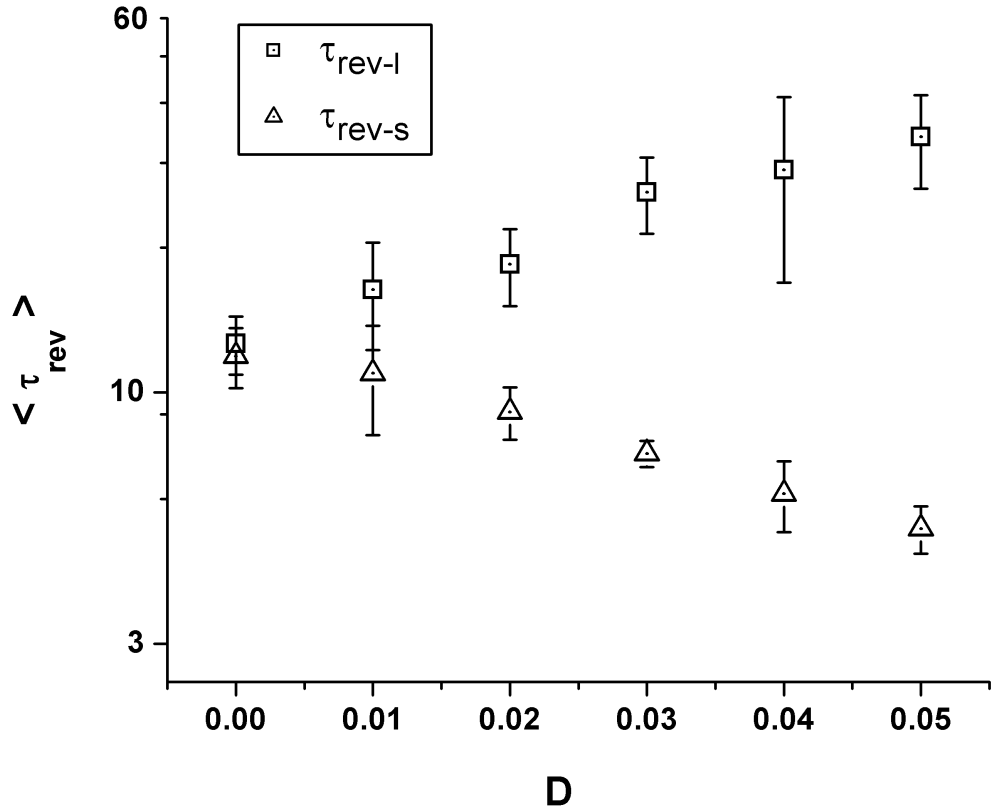,width=13cm}}
\caption{The mean reversal times (computed based on simulations of 8
track realizations with $N=1000$ motors) as a function of $D$, the
difference between the fraction of random forces along the track which
point in the favored and disfavored directions. The motion in the
favored and disfavored directions are characterized by the larger
($\tau_{\rm rev-l}$) and smaller ($\tau_{\rm rev-s}$) reversal times,
respectively.}
\label{fig:4}
\end{figure}

What happens when the simulated track is not perfectly a-polar and the
number unit cells in which the random force is pointing in one
direction is slightly larger than in the opposite direction?
Obviously, the nature of motion is expected to gradually change from
bidirectional to unidirectional. In order to investigate this
transition between two types of dynamics, we simulated the dynamics of
motors on tracks in which the fraction of random forces pointing in
one direction, $p_l$, is slightly larger than 0.5. In the simulations,
we fixed the number of motors to $N=1000$ and varied the difference
$D=p_l-(1-p_l)=2p_l-1$ between the fractions of random forces pointing
in the favored and disfavored directions.  For each track, the
simulation data was analyzed in a manner similar to that described
above for a-polar tracks, i.e., by fitting the PDF of time intervals
of unidirectional motion to an exponential function. There is one
notable difference, however, between the analysis of the results for
a-polar and for slightly polar tracks. In the latter case, two PDFs,
one corresponding for each direction of motion, must be generated with
different characteristic reversal times. The motion in the preferred
direction is characterized by the larger reversal time, $\tau_{\rm
rev-l}$, which increases with $D$. Conversely, the smaller reversal
time, $\tau_{\rm rev-s}$, corresponding to the motion in the opposite
unpreferred direction decreases with $D$. These observations are
summarized in Figs.~\ref{fig:3} and \ref{fig:4}. In Fig.~\ref{fig:3},
the PDFs corresponding to tracks with $D$ values varying from 0 to
0.05 are shown.  In the a-polar case $D=0$, the two PDFs coincide with
each others, and the velocity histogram (see inset) is bimodal. As $D$
increases, the two PDFs become increasingly different - the one
corresponding to the preferred direction of motion becomes flatter
(due to the increase in $\tau_{\rm rev-l}$), while the other one gets
more peaked at the origin (as a result of the decrease in $ \tau_{\rm
rev-s}$ ). The fact that the motors spend larger time intervals moving
in one direction is also reflected in the corresponding velocity
histograms which become less and less symmetric. The results presented
in Fig.\ref{fig:3} are obtained from simulations of six different
track realizations, one for each different value of $D$. The mean
reversal times, $\langle \tau_{\rm rev-l}\rangle$ and $\langle
\tau_{\rm rev-s}\rangle$, obtained by averaging the reversal times
computed for 8 track realizations for each value of $D$, are shown in
Fig.~\ref{fig:4}. For $D=0.05$, $\langle \tau_{\rm rev-l}\rangle\sim
10\langle \tau_{\rm rev-s}\rangle$. For even larger values of $D$, the
dynamics are essentially unidirectional, as intervals of motion in the
unpreferred direction become very rare and short. We, thus, conclude
that bidirectional motion can be observed only on a-polar or slightly
polar tracks. This conclusion is directly related to the cooperativity
of the motors which causes persistent motion that cannot be easily
reversed.

\section{Analytical treatment}
\label{sec:analytical}

\begin{figure}
  {\centering \hspace{2cm}\epsfig{file=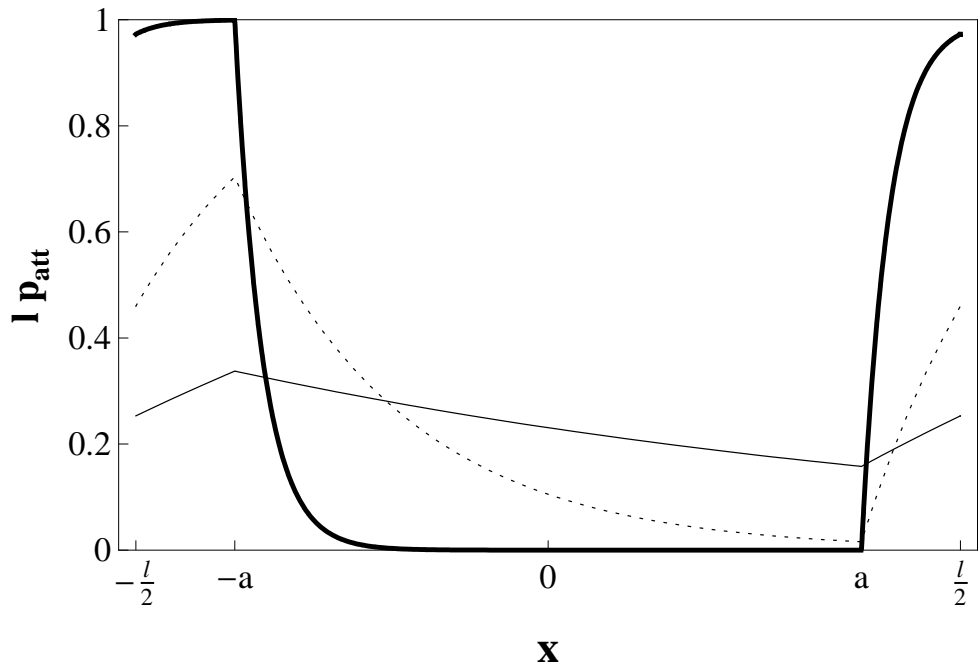,width=13cm}}
\caption{The steady state probability density, $p_{\rm att}$, as a
function of $x$, the position within a unit cell of the periodic
potential. The functions plotted in the figure correspond to
$2a=0.76l$, $\omega_3^0=0$, and $(\omega_1,\omega_2)=(v/l,v/l)$ - thin
solid line, $(\omega_1,\omega_2)=(5v/l,5v/l)$ - dashed line,
$(\omega_1,\omega_2)=(30v/l,30v/l)$ - thick solid line.}
\label{fig:5}
\end{figure}

In the following section we use mean field master equations to analyze
the bidirectional motion exhibited by our computational model. The
mean field approach corresponds to the limit $N\gg 1$ where one can
introduce the probability densities $p_{\rm att}(x)$ and $p_{\rm
det}(x)$ of finding a motor in the attached or detached state,
respectively, at position $-l/2<x\leq l/2$ within the unit cell of the
periodic potential. These probability densities are the steady-state
solutions of the following set of coupled master equations which
govern the transitions between the two connectivity states:
\begin{equation}
\left\{
\begin{array}{l}
\partial_t p_{\rm att}(x,t)+v\partial_x p_{\rm att}(x,t)
=-\omega_{\rm off}(x) p_{\rm att}(x,t)+
\omega_{\rm on} (x) p_{\rm det}(x,t)\\
\partial_t p_{\rm det}(x,t)+v\partial_x p_{\rm det}(x,t)
=-\omega_{\rm on}(x) p_{\rm det}(x,t)
+\omega _{\rm off}(x) p_{\rm att}(x,t).
\end{array}
\right.
\label{eq:master}
\end{equation}
In Eq.~\ref{eq:master}, $\omega_{\rm on}(x)$ and $\omega_{\rm off}(x)$
denote the space-dependent on and off rates, and $v$ is the group
velocity of the motors. Because the spacing between the motors is
incommensurate with the periodicity of the potential, the total
spatial distribution is uniform in $x$ for $N\gg 1$:
\begin{equation}
p_{\rm att}(x,t)+p_{\rm det}(x,t)=\frac{1}{l}.
\label{eq:uniform}
\end{equation}
Using Eq.~\ref{eq:uniform}, together with Eqs.~\ref{eq:onrate} and
\ref{eq:offrate} to define the on and off rates in
Eq.~\ref{eq:master}, the following steady-state equation ($\partial_t
p=0$) can be derived for $p_{\rm att}(x)$:
\begin{equation}
lv\frac{dp_{\rm att}(x)}{dx}=\left\{
\begin{array}{ll}
-l\omega_1p_{\rm att}(x) & \ {\rm for}\
|x|\leq a \\ 
\omega_2-l\left(\omega_2+\omega_3^0\exp\left(\alpha N\right)\right)
p_{\rm att}(x)& \ {\rm for}\ a<|x|\leq l/2.
\end{array}
\right.
\label{eq:steady-state}
\end{equation}
Eq~\ref{eq:steady-state} should be solved subject to the boundary
condition that $p_{\rm att}(-l/2)=p_{\rm att}(l/2)$ and the
requirement that $p_{\rm att}(x)$ is continuous anywhere in the
interval $-l/2\leq x\leq l/2$, including at $x=\pm a$. Several
solutions are plotted in Fig.~\ref{fig:5} for $2a=0.76l$ (see Table
\ref{tab:parameters}), $\omega_3^0=0$, and
$(\omega_{1},\omega_{2})=(v/l,v/l)$ (thin solid line), $(5v/l,5v/l)$
(dashed line), and $(30v/l,30v/l)$ (thick solid line). The solutions
correspond to the case when the motors move to the right ($v>0$) and,
therefore, it is easy to understand why $p_{\rm att}$ reaches its
maximum at $x=-a$ [just before the motors enter, from the left, into
the central gray-shaded detachment interval ($-a<x<a$)] and its
minimum at $x=a$ [just before leaving the central detachment interval
through the right side]. We also notice that when the off rate
$\omega_1\gg v/l$, $p_{\rm att}$ drops very rapidly (exponentially) to
near zero in the detachment interval. When the attachment rates
$\omega_2\gg v/l$, $p_{\rm att}$ increases exponentially fast for
$x>a$ and rapidly reaches the maximum possible value $p_{\rm
att}=1/l$. The second steady state solution corresponding to the case
when the motors move to the left ($v<0$) is simply a mirror reflection
of the first solution with respect to $x=0$.


The mean fraction of connected motors, $P$, can be obtained by
integrating the function $p_{\rm att}(x)$ over the interval $-l/2\leq
x\leq l/2$
\begin{equation}
P=\int_{-l/2}^{l/2}p_{\rm att}(x)\,dx.
\label{eq:pconnected}
\end{equation}
The population of connected motors can be divided into two groups: The
connected motors which are located left to the minimum of the periodic
potential ($-l/2<x<0$) experience forces pushing them to the right,
i.e., forces directed in their direction of motion. Conversely,
attached motors which are located right to the minimum experience
forces directed opposite to their direction of motion. Thus, the bias
parameter $\Delta$, previously defined as the excess mean fraction of
motors supporting the motion, can be related to $p_{\rm att}$ by
\begin{equation}
\Delta=\int_{-l/2}^{0}p_{\rm att}(x)\,dx-\int_{0}^{l/2}p_{\rm att}(x)\,dx.
\label{eq:biasparameter}
\end{equation}
%
In order to derive an expression for the reversal time of the
dynamics, we now consider the fluctuations of the
instantaneous bias parameter, $\Delta(t)$, around the mean value
$\Delta$. The motors may switch their direction of motion when
$\Delta(t)=0$, i.e., when the motion momentarily stops. The occurrence
probability of such an event can be related to the mean reversal by:
\begin{equation}
\tau_{\rm rev}\sim\left[\Pi\left(\Delta\left(t\right)=0\right)\right]^{-1}.
\label{eq:revtimebias}
\end{equation}
To estimate $\Pi\left(\Delta\left(t\right)=0\right)$ we proceed by
noting that the probability of finding a motor attached left to the
minimum of the potential, i.e. a motor experiencing a force directed
in the direction of motion, is $P^+=(P+\Delta)/2$. The probability that
a motor is experiencing a force directed opposite to the direction of
motion is $P^-=(P-\Delta)/2$. The probability of having $N_+$ and
$N_-\leq N_+$ motors which, respectively, support and object to the
motion can thus be approximated by the trinomial distribution function
\begin{equation}
\pi(N_+,N_-)=\frac{N!}{N_+!N_-!(N-N_+-N_-)!}\left(\frac{P+\Delta}{2}\right)
^{N_+}\left(\frac{P-\Delta}{2}\right)
^{N_-}(1-P)^{(N-N_+-N_-)}.
\label{eq:trinomial}
\end{equation}
The instantaneous bias is given by $\Delta(t)=(N_+-N_-)/N$, and the
probability that $\Delta(t)=0$ can be expressed as sum over the
relevant terms in Eq.~\ref{eq:trinomial} for which $N_-=N_+$ 
\begin{equation}
\Pi(\Delta(t)=0)=\sum_{i=0}^{N/2}\pi(i,i)=\sum_{i=0}^{N/2}
\frac{N!}{(i!)^2(N-2i)!}\
\left(\frac{P^2-\Delta^2}{4}\right)^i(1-P)^{(N-2i)}.
\label{eq:sunprob}
\end{equation}
Replacing the sum in Eq.~\ref{eq:sunprob} by an integral, using
Sterling's approximation for factorials, expanding the logarithm of
the integrand in a Taylor series (up so second order) around the
maximum which is at $i_{\rm
max}=(N/2)\sqrt{P^2-\Delta^2}/(1-P+\sqrt{P^2-\Delta^2})$ and then
exponentiating the expansion, and finally extending the limits of
integration to $\pm\infty$ (which has a negligible effect on the
result for $N\gg 1$) - leads to:
\begin{equation}
\Pi(\Delta(t)=0)=
\left[1-P+\sqrt{P^2-\Delta^2}\right]^N
\times
\int_{-\infty}^{+\infty}
dy\exp\left[-\frac{2}{C(1-C)N}\left(y-i_{\rm max}\right)^2\right],
\label{eq:taurevexp1b}
\end{equation}
where $C=\sqrt{P^2-\Delta^2}/(1-P+\sqrt{P^2-\Delta^2})$. This yields
\begin{equation}
\tau_{\rm rev}=\frac{2\tau_0}{\Pi\left(\Delta\left(t\right)=0\right)}=
2\tau_0\sqrt{\frac{2}{\pi C(1-C)N}}\left[1-P+\sqrt{P^2-\Delta^2}\right]^{-N},
\label{eq:taurevexp2}
\end{equation}
where $\tau_0$ is some microscopic time scale. (The factor of 2 in the
numerator in Eq.~\ref{eq:taurevexp2} is due to the fact that once the
motors stop, they have equal probability to move in both directions.)
As noted before (see section \ref{sec:introduction}),
Eq.~\ref{eq:taurevexp2} predicts an almost exponential dependence of
$\tau_{\rm rev}$ on $N$ only for constant values of $P$ and $\Delta$,
which was the case in ref.~\cite{badoual}. In the more general case,
the dependence of $\tau_{\rm rev}$ on $N$ can be derived by
calculating the values of $P$ and $\Delta$ as a function of $N$ and
substituting these values into Eq.~\ref{eq:taurevexp2}.

\begin{figure}
  {\centering \hspace{2cm}\epsfig{file=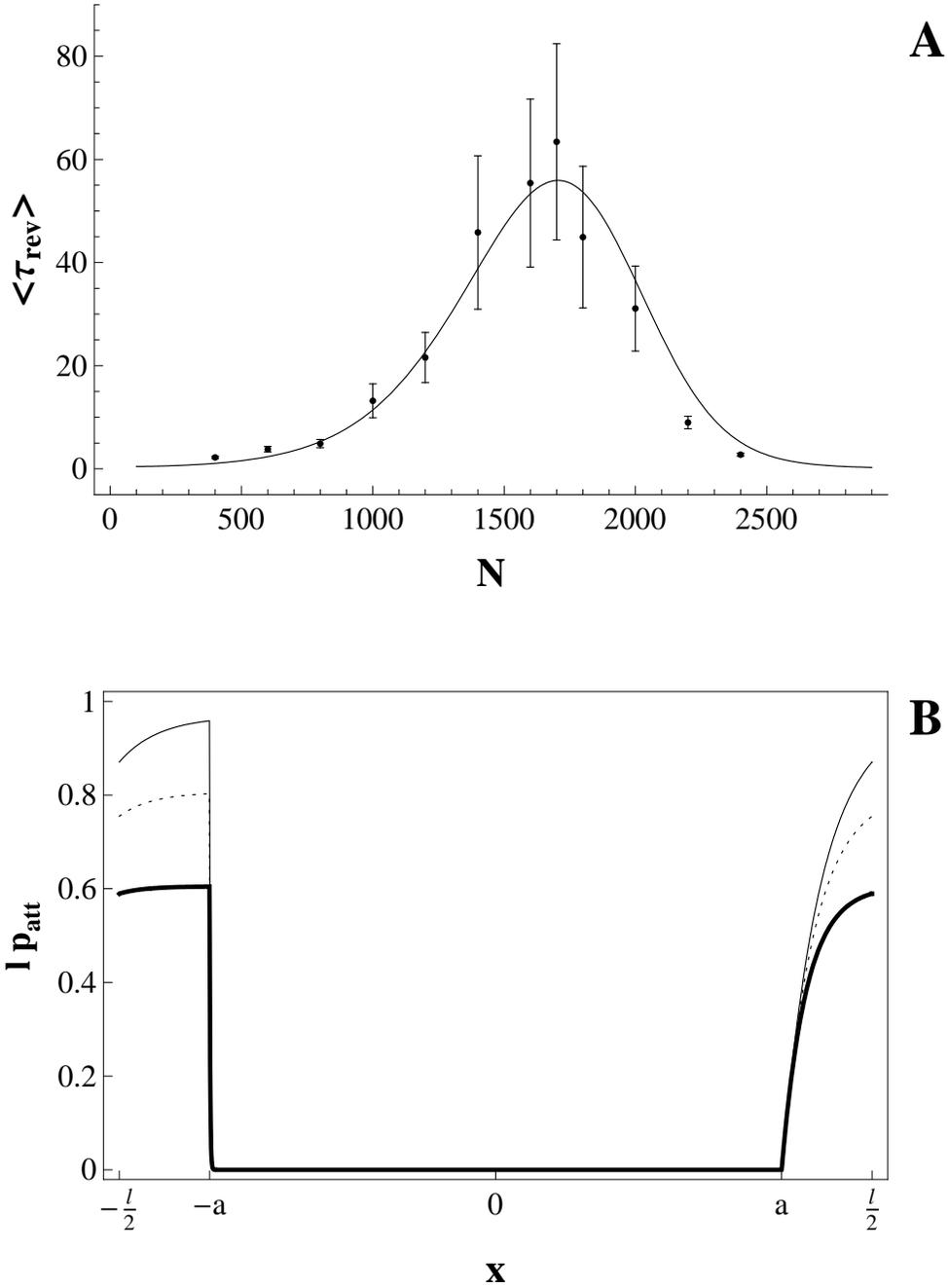,width=13cm}}
\caption{(A) The reversal time $\langle \tau_{\rm rev}\rangle$
function of the number of motors $N$. The circles denote the
simulations results (replotted from {\protect Fig.~\ref{fig:2}}). The
curve is a fit of the results to {\protect Eq.~\ref {eq:taurevexp2}},
with $\tau_0=680$ msec and $v=8.2$ nm/sec. (B) The steady state
probability density $p_{\rm att}(x)$ computed for several values of
$N$ ($N=1000$ - solid line, $N=2000$ - dashed line, $N=2500$ - thick
solid line). The group velocity of the motors is $v=8.2$ nm/sec [as in
(A), above], while the model parameters are set to the values used in
our simulations (see Table {\protect\ref{tab:parameters}}).}
\label{fig:6}
\end{figure}

To test the validity and accuracy of the analytical expression for
$\tau_{\rm rev}$, we take the following steps: (i) set the model
parameters $l$, $a$, $\omega_1$, $\omega_2$, $\omega_3^0$, and
$\alpha$ to the values used in our computer simulations which are
given in Table \ref{tab:parameters}, (ii) calculate the probability
density $p_{\rm att}$ corresponding to these values
(Eq.~\ref{eq:steady-state}) and use Eqs.~\ref{eq:pconnected} and
\ref{eq:biasparameter} to calculate $P$ and $\Delta$ over the range of
$N$ studied in the simulations, (iii) substitute the values of $P$ and
$\Delta$ into Eq.~\ref{eq:taurevexp2}, to obtain $\tau_{\rm rev}$ as a
function of $N$, (iv) fit the analytical expression for $\tau_{\rm
rev}(N)$ to the simulation results plotted in Fig.~\ref{fig:2}B. This
procedure involves two fitting parameters: the microscopic time scale
$\tau_0$ appearing in Eq.~\ref{eq:taurevexp2}, and the group velocity
$v$ appearing in the steady-state equation
(Eq.~\ref{eq:steady-state}). A seemingly reasonable choice for the
latter would be $v=20$ nm/sec, which is where the velocity histogram
of the bidirectional motion is peaked (see inset of
Fig.~\ref{fig:3}A).  However, the motors slow down before each change
in their direction of the motion; and because these changes in the
directionality are fairly rare events, their occurrence probability is
likely to be strongly influenced by the short periods of slow motion
preceding them. Thus, it can be expected that the best fit of
Eq.~\ref{eq:taurevexp2} to the simulation results is achieved for
$v<20$ nm/sec. Indeed, for $v=8.2$ nm/sec and $\tau_0=680$ msec, we
obtain the fitting curve shown in Fig.~\ref{fig:6}A, which is an
excellent agreement with our computational results for the reversal
times (plotted in Fig.~\ref{fig:2}B and replotted here in
Fig.~\ref{fig:6}A) over the whole range of values of $N$ investigated
($400<N<2400$). The steady state probability density, $p_{\rm
att}(x)$, on the basis of which $\tau_{\rm rev}$ was calculated is
shown in Fig.~\ref{fig:6}B for several different values of $N$
($N=1000$ - solid line, $N=2000$ - dashed line, $N=2500$ - thick solid
line). As can be seen from the figure, the detachment rate $\omega_1$
in our simulations is so large that the central detachment interval of
the unit cell ($-a<x<a$) is completely depleted of motors. Increasing
$N$ leads to a decrease in the effective attachment rate around the
potential maximum, which reduces both the number of motors supporting
($-l/2<x<-a$) and objecting ($a<x<l/2$) the motion and leads to the
non-monotonic dependence of $\tau_{\rm}$ on $N$. The fitting value of
$\tau_0=680$ msec is very close to $\tau^*=l/v=5\ {\rm nm}/(8\ {\rm
nm/sec})=625$ msec, which is the traveling time of the motors within a
unit cell of the potential (once we set $v=8.2$ nm/sec) and, therefore,
is also the characteristic time scale at which the motors change their
``states'' (detached, connected and supporting the motion, connected
and objecting the motion). The remarkable agreement between the
analytical and simulation results for $\tau_{\rm rev}$ should not,
however, be allowed to obscure the fact that Eq.~\ref{eq:taurevexp2}
is based on a mean field approximation which, in principle, is not
suitable for the calculating the probabilities of rare fluctuation
events (such as velocity reversals in cooperative bidirectional
movement). The agreement is achieved with effective velocity ($v=8.2$
nm/sec) which is significantly smaller than the typical velocity
measured in the simulations ($v=20$ nm/sec). Therefore, one should
not expect the steady state probability density $p_{\rm att}(x)$
plotted in Fig.~\ref{fig:6}B to perfectly match the simulations data.

\section{Summary}
\label{sec:summary}

We use a two-state ratchet model to study the cooperative
bidirectional motion of myosin II motors on actin tracks with randomly
alternating polarities. Our model is an extension of a model
previously proposed by Badoual {\em et al.}\/ to explain the
macroscopically large reversal times measured in motility assays
\cite{badoual}. These time scales of velocity reversals are orders of
magnitude longer than the microscopic typical stepping times of
individual motors and can be understood as a result of collective
effects in many-motor systems. The ratchet model that we use assumes
that the motors are coupled mechanically but act independently, i.e.,
their binding to and unbinding from the cytoskeletal track are
statistically uncorrelated. These assumptions lead to a predicted
exponential increase of $\tau_{\rm rev}$ with $N$, the number of
motors. Motivated by recent experiments which exhibit no such
dependence of $\tau_{\rm rev}$ on $N$ \cite{gilboa}, we introduced a
modified version of Badoual's model which accounts for an additional
cooperative effect of the molecular motors and which eliminates the
exponential increase of $\tau_{\rm rev}$ with $N$. This additional
collective effect arises from the forces that the motors jointly exert
on the actin and the associated elastic energy which (within a mean
field approximation) scales as $E/K_BT\sim NN_C$ (where $N_C<N$ is the
number of attached motors). This scaling relationship implies that the
typical energy released when a motor is detaching from the track
increases linearly with $N$ and, therefore, the detachment rate in
many-motor systems should be larger than the detachment rate of
individual motors. We show, both computationally and analytically,
that when this effect is taken into account and the detachment rate is
properly redefined, the characteristic reversal time does not diverge
for large $N$. Instead, $\tau_{\rm rev}$ exhibits a much weaker
dependence on $N$ and reaches a maximum at intermediate values of $N$.
 
While our model definitely improves the agreement with the experimental
results (compared to the original model), further improvement is
needed in order to eliminate the non-monotonic dependence of
$\tau_{\rm rev}$ on $N$. One step in this direction may be to consider
other forms of the off-rate $\omega_3$ which are based on more
accurate evaluations of the actin elastic energy. In the present work,
our analysis is based on a mean field approximation which makes the
calculation tractable by assuming that the detachment rate depends
only on $N$ (the total number of motors), but not on the instantaneous
number of attached motors and their locations along the cytoskeletal
track. A full statistical mechanical treatment is feasible only for
small systems, which we plan to report in a future publication. As a
final remark here we note that the mean field approximation probably
leads to over-estimation of the effect of the ``track-mediated''
elastic interactions on the reversal times (which may explain the
decrease in $\tau_{\rm rev}$ for large $N$).  In a non-mean-field
calculation the motors which release higher energy will detach at
higher rates, and the detachment of these ``energetic'' motors will
lead to the release of much of the elastic energy stored in the actin
track. By contrast, in the mean field approximation the contribution
of all the connected motors to the energy is the
same. Therefore. within the mean-field approximation, a larger number
of motors must be disconnected at a higher frequency, which increases
the ``stochastic noise'' in the system that reduced $\tau_{\rm rev}$.

We thank Haim Diamant and Yariv Kafri for useful discussions. A.B.G
wishes to thank the Joseph and May Winston Foundation Career Development
Chair in Chemical Engineering, the Israel Cancer Association (grant
No.~20070020B) and the Israel Science Foundation (grant No.~551/04).


\newpage

\end{document}